\documentclass[preprint,preprintnumbers,eqsecnum,superscriptaddress]{revtex4}
\usepackage{amsmath}
\usepackage{latexsym}
\usepackage{amsfonts}
\usepackage{graphicx}
\usepackage{bm}
\usepackage{subfigure}

\begin{document}

\preprint{KU-TP 035}

\title{Horizon Thermodynamics and Gravitational Field
Equations in Ho\v{r}ava-Lifshitz Gravity}

\author{Rong-Gen Cai}\email{cairg@itp.ac.cn}

\affiliation{Key Laboratory of Frontiers in Theoretical
Physics, Institute of Theoretical Physics, \\
Chinese Academy of Sciences, P.O. Box 2735, Beijing 100190, China }
\affiliation{Department of Physics, Kinki University, Higashi-Osaka,
Osaka 577-8502, Japan}

\author{Nobuyoshi Ohta}\email{ohtan@phys.kindai.ac.jp}
\affiliation{Department of Physics, Kinki University, Higashi-Osaka,
Osaka 577-8502, Japan}

\begin{abstract}
We explore the relationship between the first law of thermodynamics and
gravitational field equation at a static, spherically symmetric
black hole horizon in Ho\v{r}ava-Lifshtiz theory with/without
detailed balance.  It turns out that as in the cases of Einstein
gravity and Lovelock gravity, the gravitational field equation can
be cast to a form of the first law of thermodynamics at the black
hole horizon.  This way we obtain the expressions for entropy and
mass in terms of black hole horizon, consistent with those from
other approaches. We also define a generalized Misner-Sharp energy
for static, spherically symmetric spacetimes in Ho\v{r}ava-Lifshtiz
theory. The generalized Misner-Sharp energy is conserved in the case
without matter field, and its variation gives the first law of black
hole thermodynamics at the black hole horizon.

\end{abstract}

\maketitle

\section{Introduction}

The holographic principle might be one of
the principles of nature, which states that a theory with gravity could
be equivalent to a theory without gravity in one less dimension.
The well-known AdS/CFT correspondence~\cite{Mald} is a realization
of the holographic principle, while the latter is motivated by black
hole thermodynamics. The black hole thermodynamics says that a black
hole behaves as an ordinary thermodynamic system with temperature
and entropy. The temperature of a black hole is proportional to
surface gravity at its horizon, while the entropy of the black hole
is measured by its horizon area. Black hole mass, temperature and
entropy satisfy the first law of thermodynamics. These results come
from a combination of quantum mechanics, black hole geometry and
general relativity. This implies that there might exist a deep
connection between thermodynamics and gravity theory.

Indeed some pieces of evidence have been accumulated for the
connection between thermodynamics and gravity theory in the
literature. Assuming there is a proportionality between entropy
and horizon area, Jacobson~\cite{Jac} derived the Einstein field
equation by using the fundamental Clausius relation, $\delta Q=
TdS$, connecting heat, temperature and entropy. The key idea is to
demand that this relation holds for all the local Rindler causal
horizon  through each spacetime point, with $\delta Q$ and $T$
interpreted as the energy flux and Unruh temperature seen by an
accelerated observer just inside the horizon. In this way,
the Einstein field equation is nothing but an equation of state of
spacetime.  More recently, Jacobson's argument has been
generalized to all diffeomorphism-invariant theories of
gravity~\cite{Brus}(however, see also~\cite{Parikh}). For $ f(R)$
theory and scalar-tensor theory, see also \cite{ES,WuYZ}. In fact,
investigating the thermodynamics of spacetime for  $f(R)$
theory~\cite{Jac1,AC1,AC} and scalar-tensor theory~\cite{AC1,CC1},
it is found that a nonequilibrium thermodynamic setup has to be
employed. Further, it is argued that if shear of spacetime is not
assumed to vanish, the nonequilibrium thermodynamic setting is
required even for the Einstein general relativity~\cite{Eling,CL}.
There an internal entropy production
term has to be introduced to balance energy conservation. The
internal entropy production term $dS_i$ is proportional to the
squared shear of the horizon and the ratio of the proportionality
constant to the area entropy density is $1/4\pi$. The latter is a
universal value for many kinds of conformal field theories with
AdS duals~\cite{KSS}.

There exists another route in  exploring the relationship
between thermodynamics and gravity theory. Padmanabhan~\cite{Pad1}
first noticed that the gravitational field equation in a static,
spherically symmetric spacetime can be rewritten as a form of the
ordinary first law of thermodynamics at a black hole horizon. This
indicates that Einstein's equation is nothing but a thermodynamic
identity. For a recent review on this, see \cite{Pad2}.  This
observation was then extended to the cases of stationary
axisymmetric horizons and evolving spherically symmetric horizons
in the Einstein gravity~\cite{KSP}, static spherically symmetric
horizons~\cite{KP} and dynamical apparent horizons~\cite{CCHK} in
Lovelock gravity, and three-dimensional Banados-Teitelboim-Zanelli black hole
horizons~\cite{Akbar1}. On the other hand, the relationship
between the first law of thermodynamics and dynamical equation of
spacetime has been intensively investigated in a Friedmann-Robertson-Walker (FRW)
cosmological setup in various gravity
theories~\cite{CK,others,AC07,CC2,AC1,AC,CC1,GW,Others1}; it is
shown that (modified) Friedmann equations can be cast to a form of
the first law of thermodynamics, and there exists a Hawking
radiation associated with apparent horizon in a FRW
universe~\cite{CCH}.

Recently a field theory model for a UV complete theory of gravity
was proposed by Ho\v{r}ava~\cite{Hor}, which is a nonrelativistic
renormalizable theory of gravity and is expected to recover
Einstein's general relativity at large scales. This theory is
named Ho\v{r}ava-Lifshitz theory in the literature since at the UV
fixed point of the theory space and time have different scalings.
Since then a lot of work has been done in exploring various
aspects of the theory; for a more or less complete list of
references, see, for example, \cite{WDR}.

In this paper we discuss the relationship between the first law of
thermodynamics and gravitational field equation in
Ho\v{r}ava-Lifshitz theory. In static spherically symmetric black
hole spacetimes, we show that the gravitational field equation can
be rewritten as $dE-TdS =PdV$ at the black hole horizon. Note that
in Ho\v{r}ava-Lifshitz theory the full diffeomorphism invariance
is broken to the ``foliation-preserving" diffeomorphism. Therefore
our result is a nontrivial generalization of Padmanabhan's
observation. In addition, we discuss the question of whether one can
define a generalized Misner-Sharp quasilocal energy in
Ho\v{r}ava-Lifshitz theory. The answer is positive.  We define a
generalized Misner-Sharp energy. It is a conserved charge when
the matter field is absent, while its variation at a black hole
horizon gives the first law of black hole thermodynamics.

This paper is organized as follows. In the next section we review
the Padmanabhan's  observation by extending his discussion to a more
general spherically symmetric spacetime.  In Sec.~III  we consider
black hole spacetimes in Ho\v{r}ava-Lifshitz theory. In Sec.~IV the
case of IR modified Ho\v{r}ava-Lifshitz theory is discussed. In
Sec.~V we define a generalized Misner-Sharp quasilocal energy for
static, spherically symmetric spacetimes in Ho\v{r}ava-Lifshitz
theory and discuss its properties. The conclusion is given in
Sec.~VI.

\section{Black Holes in Einstein Gravity}

As a warm-up exercise, in this section, we will briefly review the
observation made by Padmanabhan~\cite{Pad1} by generalizing his
discussion to a more general spherically symmetric case. In Einstein's general
relativity, the gravitational field equations are
\begin{equation}
\label{2eq1}
G_{\mu\nu}= R_{\mu\nu}-\frac{1}{2} Rg_{\mu\nu}=8\pi G T_{\mu\nu},
\end{equation}
where $G_{\mu\nu}$ is Einstein tensor and $T_{\mu\nu}$ is the energy-momentum
tensor of matter field. On the other hand, for a  general static, spherically
symmetric spacetime, its metric can be written down as
\begin{equation}
\label{2eq2}
ds^2 =- f(r) dt^2 +f^{-1}(r)dr^2 + b^2(r)(d\theta^2
 +\sin^2\theta d\varphi^2),
\end{equation}
where $f(r)$ and $b(r)$ are two functions of the radius coordinate
$r$. (Note that in the Padmanabhan's discussion~\cite{Pad1}, the
metric is assumed in a form (\ref{2eq2}) with $b(r)=r$; when
matter is present, however, such a metric form is not always
satisfied. See also~\cite{KSP}.)  Suppose the metric (\ref{2eq2})
describes a nonextremal black hole with horizon at $r_+$, then the
function $f(r)$ has a simple zero at $r=r_+$. Namely,
$f'(r)|_{r=r_+}=0$, but $f''(r)|_{r=r_+} \ne 0$. It is easy to
show that the Hawking temperature of the black hole associated
with the horizon $r_+$ is
\begin{equation}
\label{2eq3}
T = \frac{1}{4\pi} f'(r)|_{r=r_+} \equiv \frac{1}{4\pi} f'(r_+),
\end{equation}
where a prime stands for the derivative with respective to $r$.
 Einstein's equations in the metric (\ref{2eq2}) have the
components
\begin{eqnarray}
\label{2eq4} G^t_t &=& \frac{1}{b^2} (-1 +f b'^2 +b (f'b'+2fb'')),
\nonumber \\
G^r_r &=& \frac{1}{b^2}(-1+b f'b'+fb'^2).
\end{eqnarray}
Note that at the horizon, one has $f(r)=0$, and then
\begin{equation}
\label{2eq5}
G^t_t|_{r=r_+}=G^r_r|_{r=r_+}=\frac{1}{b^2}(-1 +b f' b')|_{r=r_+}.
\end{equation}
Therefore at the horizon, the $t-t$ component of Einstein's
equations can be expressed as
\begin{equation}
\label{2eq6}
-1 + bf'b' = 8\pi G b^2 P,
\end{equation}
where $P=T^r_r|_{r=r_+}$ is the radial pressure of matter at the horizon.
Note that here (\ref{2eq5}) guarantees
$T^t_t=T^r_r$  at the horizon. Now we multiply  $dr_+$ on both
sides of (\ref{2eq6}) and rewrite this equation as
\begin{equation}
\label{2eq7}
\frac{1}{2G} bf'b' dr_+-\frac{1}{2G}dr_+= 4\pi b^2 P dr_+.
\end{equation}
Note that $b$ is a function of $r$ only and $f'$ has a relation
to the Hawking temperature as (\ref{2eq3}). One then can rewrite
the above equation as
\begin{equation}
\label{2eq8}
T d\left (\frac{4\pi b^2}{4G}\right)-d\left(\frac{r_+}{2G}\right)
= P dV,
\end{equation}
where $dV= 4\pi b^2 dr_+$. Therefore $V$ is just the volume of
the black hole with horizon radius $r_+$ in the coordinate (\ref{2eq2}).
The equation
(\ref{2eq8}) can be further rewritten as
\begin{equation}
\label{2eq9}
TdS -dE =PdV,
\end{equation}
with identifications
\begin{equation}
S = \frac{4\pi b^2}{4G}= \frac{A}{4G}, \ \ \ E= \frac{r_+}{2G}.
\end{equation}
Clearly here $S$ is precisely the entropy of the black hole,
while $E$ is the Misner-Sharp energy~\cite{MS} at the horizon. Thus we have
shown in general that at black hole horizon, Einstein's equations can be cast
into the form of the first law of thermodynamics.

\section{Black Holes in Ho\v{r}ava-Lifshitz Gravity}

In the $(3+1)$-dimensional Arnowitt-Deser-Misner formalism, where the metric can be
written as
\begin{equation}
\label{3eq1} ds^2=-N^2 dt^2 +g_{ij}(dx^i+N^idt)(dx^j+N^jdt),
\end{equation}
and for a spacelike hypersurface with a fixed time, its extrinsic
curvature $K_{ij}$ is
\begin{equation}
\label{3eq2} K_{ij}=\frac{1}{2N}(\dot
g_{ij}-\nabla_iN_j-\nabla_jN_i),
\end{equation}
where a dot denotes a derivative with respect to $t$ and covariant
derivatives defined with respect to the spatial metric $g_{ij}$.

The action of  Ho\v{r}ava-Lifshitz theory is~\cite{Hor,LMP}
\begin{eqnarray}
\label{3eq3}
I &=& \int dtd^3x ({\cal L}_0 +{\cal L}_1 +{\cal L}_m), \\
 {\cal L}_0 &=& \sqrt{g}N \left \{\frac{2}{\kappa^2}
(K_{ij}K^{ij}-\lambda K^2) +\frac{\kappa^2\mu^2 (\Lambda
R-3\Lambda^2)}{8(1-3\lambda)}\right \},  \nonumber \\
 {\cal L}_1  &=& \sqrt{g}N \left \{\frac{\kappa^2\mu^2(1-4\lambda)}{32(1-3\lambda)}R^2
-\frac{\kappa^2}{2\omega^4}\left(C_{ij}-\frac{\mu
\omega^2}{2}R_{ij}\right)
\left(C^{ij}-\frac{\mu\omega^2}{2}R^{ij}\right) \right\}.\nonumber
\end{eqnarray}
where $\kappa^2$, $\lambda$, $\mu$, $\omega$ and $\Lambda$ are
constant parameters and the Cotten tensor, $C_{ij}$, is defined by
\begin{equation}
\label{3eq4} C^{ij}=\epsilon^{ikl} \nabla_k \left (R^j_{\
l}-\frac{1}{4}R\delta^j_l\right) = \epsilon^{ikl}\nabla_k R^j_{\
l} -\frac{1}{4}\epsilon^{ikj}\partial_kR.
\end{equation}
 The first two terms in ${\cal L}_0$ are the kinetic terms,
others in $({\cal L}_0+{\cal L}_1)$ give the potential of the
theory in the so-called ``detailed-balance" form, and ${\cal L}_m$
stands for the Lagrangian of other matter field.

Comparing the action to that of general relativity, one can see
that the speed of light, Newton's constant and the cosmological
constant are
\begin{equation}
\label{3eq5}
c=\frac{\kappa^2\mu}{4}\sqrt{\frac{\Lambda}{1-3\lambda}}, \ \
G=\frac{\kappa^2 c}{32\pi}, \ \ \tilde
\Lambda=\frac{3}{2}\Lambda,
\end{equation}
respectively. Let us notice that when $\lambda=1$,  ${\cal L}_0$
could be reduced to the usual Lagrangian  of Einstein's general
relativity. Therefore it is expected that general relativity could
be approximately recovered at large distances when $\lambda=1$.
Here we will mainly consider the case of $\lambda=1$, but will also discuss
the $\lambda\neq 1$ case briefly at the end of this paper.

Now we consider black hole spacetime with metric
ansatz~\cite{LMP,CCO}
\begin{equation}
\label{3eq6}
ds^2 =-\tilde N^2(r)f(r) dt^2 +\frac{dr^2}{f(r)} +r^2
d\Omega_k^2,
\end{equation}
where $d\Omega_k^2$ denotes the line element for a two-dimensional
Einstein space with constant scalar curvature $2k$. Without loss
of generality, one may take $k=0$, $\pm 1$, respectively.
Substituting the metric (\ref{3eq6}) into (\ref{3eq3}), we have
\begin{eqnarray}
\label{3eq7}
I & =& \frac{\kappa^2\mu^2 \Lambda
\Omega_k}{8(1-3\lambda)}\int dt dr \tilde N \left \{-3\Lambda r^2
-2 (f-k) -2 r (f-k)' \right.
\nonumber \\
&&~~~~~~~~~~ \left. +\frac{(\lambda-1)f'^2}{2\Lambda}
+\frac{(2\lambda-1)(f-k)^2}{\Lambda r^2} -\frac{2\lambda
 (f-k)}{\Lambda r}f' + \alpha r^2 {\cal L}_m \right \},
\end{eqnarray}
where a prime denotes the derivative with respect to $r$,
$\Omega_k$ is the volume of the two-dimensional Einstein space and
the constant $\alpha = 8 (1-3\lambda)/\kappa^2\mu^2 \Lambda$. In
the case of $\lambda=1$ we can rewrite the action as
\begin{equation}
\label{3eq8}
I = \frac{\kappa^2\mu^2
\sqrt{-\Lambda}\Omega_k}{16}\int dt dx \tilde N \left \{\left(
x^3-2x (f-k)+\frac{(f-k)^2}{x}\right)'+
x^2(\frac{\alpha}{-\Lambda})
 {\cal L}_m \right\}.
\end{equation}
Note that  here $x=\sqrt{-\Lambda}r$, a prime becomes the
derivative with respect to $x$. Varying the action with $ \tilde N
$, we obtain the equations of motion
\begin{equation}
\label{3eq9}
 -\frac{\kappa^2\mu^2
\sqrt{-\Lambda}\Omega_k}{16}\left ( 3x^2-2 (f-k)-
\frac{(f-k)^2}{x^2}-2x f'+\frac{2(f-k)f'}{x} \right ) = x^2
\frac{\Omega_k}{(-\Lambda)^{3/2}}\frac{\delta (\tilde N {\cal
L}_m)}{\delta \tilde N}.
\end{equation}
Suppose the nonextremal black hole (\ref{3eq6}) has a  horizon
radius $r_+$, namely $x_+= \sqrt{-\Lambda}r_+$. Then the Hawking
temperature of the black hole is
\begin{equation}
\label{3eq10}
 T = \frac{1}{4\pi} \tilde N(r)\left.
\frac{df}{dr}\right |_{r=r_+}= \frac{\sqrt{-\Lambda}}{4\pi}\tilde
N(x)f'|_{x=x_+}.
\end{equation}
Now we consider a class of solutions with $\tilde N(r)=$ const.
For example, the charged black hole solution discussed in \cite{CCO}
belongs to this class of solutions.  In this case one can set
$\tilde N=1$ by rescaling the time coordinate $t$. Note that here
not all solutions with matter field have the form $\tilde N=1$. At
the horizon $x_+$, Eq.~(\ref{3eq9}) is reduced to
\begin{equation}
 -\frac{\kappa^2\mu^2
\sqrt{-\Lambda}\Omega_k}{16}\left ( 3x^2_++2k-
\frac{k^2}{x^2_+}-2x_+ f'-\frac{2kf'}{x_+} \right ) = x^2_+
\frac{\Omega_k}{(-\Lambda)^{3/2}}\left.\frac{\delta (\tilde N
{\cal L}_m)}{\delta \tilde N}\right|_{x=x_+}.
\end{equation}
Multiplying both sides with $dx_+$, a variation of the horizon
radius, we have
\begin{equation}
 \frac{\kappa^2\mu^2
\sqrt{-\Lambda}\Omega_k}{16}\left ( 2(x_++\frac{k}{x_+})f' dx_+ -
(3x^2_++2k- \frac{k^2}{x^2_+})dx_+ \right )= x^2_+
\frac{\Omega_k}{(-\Lambda)^{3/2}}P dx_+,
\end{equation}
where $P= \left.\frac{\delta (\tilde N {\cal L}_m)}{\delta \tilde
N}\right|_{x=x_+}$.  Note that the Hawking temperature turns to be
$T= f'\sqrt{-\Lambda}/4\pi$ when $\tilde N=1$. The above equation
then can be rewritten as
\begin{equation}
\label{3eq13}
 TdS -dE = PdV,
\end{equation}
where
\begin{eqnarray}
\label{3eq14}
 S &=& \frac{\pi \kappa^2 \mu^2
\Omega_k}{4}\left(x_+^2 + 2k \ln x_+ \right)+S_0,
\nonumber \\
E &=&\frac{\kappa^2\mu^2 \sqrt{-\Lambda}\Omega_k}{16x_+}
\left(x_+^2 +k \right)^2,
\end{eqnarray}
$V= \frac{\Omega_k}{3}r_+^3$, and $S_0$ is an undetermined constant. Clearly $V$
is the volume of black hole with radius $r_+$. Comparing (\ref{3eq14}) with black
hole entropy and mass defined through a Hamiltonian approach in our previous
papers \cite{CCO}, we see
that  $S$ and $E$ are just black hole entropy and mass in terms of horizon radius $x_+$,
and the gravitational field equation at the black hole horizon
can be cast to the form of the first law of thermodynamics. Note
that here we have obtained expressions for black hole entropy and
mass, but have not used any explicit black hole solutions. In
other words, the above way provides a universal method to derive black
hole entropy and mass.

Now we turn to the case without the detailed-balance condition by
considering the action as~\cite{LMP,CCO}
\begin{equation}
\label{3eq15} I = \int dtd^3x ({\cal L}_0 + (1-\epsilon^2) {\cal
L}_1 + {\cal L}_m)
\end{equation}
where the parameter $\epsilon^2 \ne 0$. In this case, instead of
(\ref{3eq8}), we have
\begin{equation}
\label{2eq17} I = \frac{\kappa^2\mu^2
\sqrt{-\Lambda}\Omega_k}{16}\int dt dx \tilde N \left \{\left(
x^3-2x (f-k)+(1-\epsilon^2)\frac{(f-k)^2}{x}\right)'+x^2(\frac{\alpha}{-\Lambda})
 {\cal L}_m  \right\}.
\end{equation}
Varying the action with respect to $\tilde N$ yields
 \begin{eqnarray}
&&  - \frac{\kappa^2\mu^2 \sqrt{-\Lambda}\Omega_k}{16}\left (
3x^2-2 (f-k)-(1-\epsilon^2)
\frac{(f-k)^2}{x^2}\right. \nonumber \\
 &&~~~~~~~~~~~~~~~~- \left. 2x
f'+(1-\epsilon^2)\frac{2(f-k)f'}{x} \right ) = x^2
\frac{\Omega_k}{(-\Lambda)^{3/2}}\frac{\delta (\tilde N {\cal
L}_m)}{\delta \tilde N} .
\end{eqnarray}
At the black hole horizon where $f=0$, the equation reduces to
\begin{equation}
\label{3eq18}
 -\frac{\kappa^2\mu^2
\sqrt{-\Lambda}\Omega_k}{16}\left ( 3x^2_++2k-(1-\epsilon^2)
\frac{k^2}{x^2_+}-2x_+ f'-(1-\epsilon^2)\frac{2kf'}{x_+} \right )
= x^2_+ \frac{\Omega_k}{(-\Lambda)^{3/2}}\left.\frac{\delta
(\tilde N {\cal L}_m)}{\delta \tilde N}\right|_{x=x_+}.
\end{equation}
Multiplying $dx_+$ on both sides, one can express this equation as
the form (\ref{3eq13}) with the condition $\tilde N=1$, again. But
this time we have
\begin{eqnarray}
 S &=& \frac{\pi \kappa^2 \mu^2
\Omega_k}{4}\left(x_+^2 + 2k(1-\epsilon^2) \ln x_+ \right)+S_0,
\nonumber \\
E &=&\frac{\kappa^2\mu^2 \sqrt{-\Lambda}\Omega_k}{16x_+}
\left(x_+^4+2kx_+ +(1-\epsilon^2)k^2 \right).
\end{eqnarray}
These are nothing but the entropy and mass, expressed in terms of horizon
radius $ x_+$,  of the black hole  solutions found in \cite{CCO}.

Now we turn to the case with $z=4$ terms, where $z$ is the dynamical
critical exponent.  Such terms are super-renormalizable ones. The
vacuum black hole solution for this case has been discussed in
\cite{CLS}. Including $z=4$ terms changes ${\cal L}_1$ in
(\ref{3eq3}) to
\begin{eqnarray}
{\cal{L}}_1
&=&-\sqrt{g}N\frac{\kappa^2}{8}\Big\{\frac{4}{\omega^4}C^{ij}C_{ij}
-\frac{4\mu}{\omega^2}C^{ij}R_{ij}
-\frac{4}{\omega^2M}C^{ij}L_{ij}+\mu^2G_{ij}G^{ij}+\frac{2\mu}{M}G^{ij}L_{ij} \nonumber\\
&&~ +\frac{2\mu}{M}\Lambda
L+\frac{1}{M^2}L^{ij}L_{ij}-\tilde{\lambda}\big(\frac{L^2}{M^2}-\frac{\mu
L}{M}(R-6\Lambda)+\frac{\mu^2}{4}R^2\big)\Big\},
\end{eqnarray}
where
\begin{eqnarray}
G^{ij}&=&R^{ij}-\frac{1}{2}g^{ij}R\nonumber\\
L^{ij}&=&(1+2\beta)(g^{ij}\nabla^2-\nabla^i\nabla^j)R
+\nabla^2G^{ij}\nonumber\\&&~~+2\beta
R(R^{ij}-\frac{1}{4}g^{ij}R)+
2(R^{imjn}-\frac{1}{4}g^{ij}R^{mn})R_{mn}, \nonumber \\
 L &\equiv &
g^{ij}L_{ij}
=\bigg(\frac{3}{2}+4\beta\bigg)\nabla^2R+\frac{\beta}{2}R^2+\frac{1}{2}R_{ij}R^{ij},
\end{eqnarray}
and $\tilde \lambda= \lambda/(3\lambda-1)$, $\beta$ and $M$ are
two new parameters. When $\beta =-3/8$ and $\lambda=1$, the action
in the metric (\ref{3eq6}) reduces to
\begin{eqnarray}
\label{3eq22}
I&=&\frac{{\kappa}^2\Omega_k}{16\sqrt{-\Lambda^3}}\int dt dx
\tilde{N}\left \{\Big [\tilde{\mu}^2\Big
(x^3-2x(f-k)+\frac{(f-k)^2}{x}\Big )\right.
\nonumber\\
&& \left. ~~-2\tilde{\beta}\tilde{\mu}\Big (\frac{(f-k)^3}{x^3}
-\frac{(f-k)^2}{x}\Big )+\tilde{\beta}^2\frac{(f-k)^4}{x^5}\Big
]^{'} +x^2\tilde \alpha {\cal L}_m \right\},
\end{eqnarray}
where we define $\tilde{\mu}=-\mu \Lambda,
\tilde{\beta}=\frac{\Lambda^2}{4M}$, $\tilde \alpha = 16/\kappa^2
$, and the prime is still the derivative with respect to $x$.
Varying the action with respect to $\tilde N$ yields
\begin{eqnarray}
&& -\frac{{\kappa}^2 \mu^2\sqrt{-\Lambda}\Omega_k}{16}
\Big(x^3-2x(f-k)+\frac{(f-k)^2}{x}
\nonumber\\
&&~~~~~~-2\frac{\tilde{\beta}}{\tilde{\mu}}(\frac{(f-k)^3}{x^3}
-\frac{(f-k)^2}{x})+\frac{\tilde{\beta}^2}{\tilde{\mu}^2}\frac{(f-k)^4}{x^5}\Big)^{'}
=x^2 \frac{\Omega_k}{(-\Lambda)^{3/2}}\frac{\delta (\tilde N {\cal
L}_m)}{\delta \tilde N}
\end{eqnarray}
Again, we take the values of all quantities at the black hole
horizon and then multiply $dx_+$ on both sides, the above equation
turns to be
\begin{equation}
TdS-dE = PdV,
\end{equation}
where $P=\left. \frac{\delta (\tilde N {\cal L}_m)}{\delta \tilde
N}\right|_{x=x_+}$, $V= \Omega_k r_+^3/3$ is the volume of the
black hole, and
\begin{eqnarray}
E &=& \frac{\kappa^2  \mu^2 \sqrt{-\Lambda}\Omega_k}{16} \left (
x^3_+ +2kx_++\frac{k^2}{x_+} +2\frac{\tilde \beta}{\tilde
\mu}(\frac{k^3}{x_+^3} +\frac{k^2}{x}) +\frac{\tilde
\beta^2}{\tilde \mu^2}\frac{k^4}{x_+^5}\right)
\nonumber \\
S &=& \frac{\pi \kappa^2 \mu^2 \Omega_k}{4} \left ( x_+^2 +2k\ln
x_+ -3\frac{\tilde \beta k^2}{\tilde \mu x_+^2}
 -\frac{\tilde \beta^2k^3}{\tilde \mu^2 x_+^4} +
4\frac{\tilde \beta k}{\tilde \mu} \ln x_+ \right) +S_0
\end{eqnarray}
This way we have obtained  entropy and mass of black hole
solutions~\cite{CLS}, again, and shown that at the black hole horizon,
gravitational field equation can be cast into the form of the first law of
thermodynamics.

\section{Black Holes in IR Modified Ho\v{r}ava-Lifshitz Gravity}

In this section we consider the case with broken detailed-balance
by introducing a term $\mu^4 R$ to the action~\cite{KS}. Such
theory is called IR modified Ho\v{r}ava-Lifshitz theory.  In this
way, it is found that one can get asymptotically flat solutions.
In fact, introducing the parameter $\epsilon^2$ to the original action
of Ho\v{r}ava-Lifshitz theory with detailed-balance can also lead
to asymptotically flat solutions~\cite{LMP,CCO}. Here we show
that for the IR modified Ho\v{r}ava-Lifshitz theory, gravitational
field equations at the black hole horizon can also be cast into a form of
the first law of thermodynamics.

Now we add a new term
\begin{equation}
\label{4eq1}
 {\cal L}_3 = \sqrt{g} N\frac{\kappa^2 \mu^2
\nu}{8(3\lambda -1)} R,
\end{equation}
to the action (\ref{3eq3}). Here $\nu$ is a new parameter. The
term (\ref{4eq1}) ``softly" violates the so-called detailed-balance.
The action in the metric (\ref{3eq6}) changes
to~\cite{Park}
\begin{eqnarray}
\label{4eq2}
 I  &=& \frac{\kappa^2\mu^2 \Omega}{8(1-3\lambda)}
\int dt d r
\tilde N \left ( (2\lambda-1) \frac{(f-1)^2}{r^2} - 2\lambda
\frac{f-1}{r}f' + \frac{\lambda-1}{2}f'^2 \right. \nonumber \\
&&~~~~~~~~~~- 2(\nu -\Lambda)(1-f-rf')-3\Lambda^2 r^2 + \alpha \Lambda r^2 {\cal L}_m ),
\end{eqnarray}
where we have restricted to the case with $k=1$ and a prime stands
for the derivative with respect to $r$. Considering the case with
$\lambda=1$, and varying the action with respect to $\tilde N$, one
has the equation of motion
\begin{equation}
\frac{\kappa^2\mu^2 \Omega}{16} \left ( \frac{(f-1)^2}{r^2} - 2
\frac{f-1}{r}f'- 2(\nu -\Lambda)(1-f-rf')-3\Lambda^2 r^2 \right ) = r^2
\Omega
\frac{\delta (\tilde N {\cal L}_m)}{\delta \tilde N},
\end{equation}
At a black hole horizon where $f=0$ and $f'|_{r=r_+}= 4\pi T$,
by the same approach, we can rewrite the equation as
\begin{equation}
TdS -dE =PdV,
\end{equation}
where
\begin{eqnarray}
S &=& \frac{\pi \kappa^2 \mu^2 \Omega}{4}\left((\nu -\Lambda)
r_+^2 +2 \ln r_+\right) +S_0,
\nonumber \\
E &=&\frac{\kappa^2\mu^2\Omega}{16} \left( \Lambda^2 r_+^3
+2(\nu-\Lambda)r_+ +\frac{1}{r_+}\right).
\end{eqnarray}
This energy is the same as that given in \cite{Park}, up to a
factor, which is not figured out there. The entropy is given
for the first time, although related results on thermodynamics of black holes
in the modified Ho\v{r}ava-Lifshitz theory have been discussed in
\cite{Myung}.

\section{Generalized Misner-Sharp energy and the first law of black
hole thermodynamics}

Quasilocal energy is an important concept in general relativity.
In particular, the so-called Misner-Sharp energy is intensively
discussed in the literature. In a spherically symmetric spacetime
with metric
\begin{equation}
ds^2 = h_{ab}dx^adx^b + r^2(x) (d\theta^2 +\sin^2\theta
d\varphi^2),
\end{equation}
where $a=0$ and $1$, the Misner-Sharp energy is defined
as~\cite{MS}
\begin{equation}
\label{5eq2}
 M(r)=\frac{r}{2G}\left(1-h^{ab}\partial_ar \partial_b
r\right),
\end{equation}
which is valid for general relativity in four dimensions. For
Schwarzschild solution, (\ref{5eq2}) just gives the Schwarzschild
mass, while it gives the effective Schwarzschild mass $m(r)$ at
$r$ for a static spherically symmetric spacetime (\ref{3eq6}) with
$f(r)= 1-\frac{2Gm(r)}{r}$. Therefore at a black hole horizon
$r_+$, the Misner-Sharp energy (\ref{5eq2}) gives us the
energy of gravitational field at the horizon $r_+$: $M=r_+/2G$.

In the previous sections, we have shown that the gravitational
field equations at a black hole horizon can be cast to a form of
the first law of thermodynamics in Ho\v{r}ava-Lifshtiz theory. In
this section, we show that the form of action for the Ho\v{r}ava-Lifshtiz
theory allows us to give a generalized Misner-Sharp quasilocal
energy in the case of static, spherically symmetric spacetime
(\ref{3eq6}).

Let us start with the action (\ref{3eq3}) with the detailed-balance.
In this case, the gravitational part of the action can be
rewritten in a derivative form (\ref{3eq8}), which enables us to
define a generalized Misner-Sharp energy as
\begin{equation}
\label{5eq3}
M(r) =\frac{\kappa^2\mu^2
\Omega_k}{16\ r}\left ( \Lambda^2 r^4-2 \Lambda r^2 (k-f)+ (k-f)^2
\right ).
\end{equation}
It is easy to see that at a black hole horizon $r_+$, this
quasilocal energy $E(r)$ gives the mass (\ref{3eq14}) of the black
hole solution. The variation of the generalized Misner-Sharp
energy with respect to $r$ gives
\begin{equation}
\label{5eq4}
dM (r) = -r^2 \Omega_k \frac{\delta (\tilde N {\cal
L}_m)}{\delta \tilde N} dr,
\end{equation}
from which one can see clearly that $-\frac{\delta (\tilde N {\cal
L}_m)}{\delta \tilde N}$ is the energy density of matter field. In
the case without matter field, the generalized Misner-Sharp energy
is conserved, $dM(r)=0$. At the horizon we have
\begin{equation}
\label{5eq5}
dM(r)|_{r=r_+} = dE -T dS,
\end{equation}
where $E$ and $S$ are just mass and entropy of the black hole,
given in (\ref{3eq14}). When the matter field is absent, it gives
us $dE =TdS$, which is the first law of black hole
thermodynamics. In \cite{CCO}, we have used the first law to
derive the black entropy.

In the case including the $z=4$ term, the generalized Misner-Sharp
energy can be read down from the action (\ref{3eq22})
\begin{eqnarray}
M(r) &=& \frac{\kappa^2\mu^2\Omega_k}{16 \ r}
 \left ( \Lambda^2 r^4-2 \Lambda r^2 (k-f)+ (k-f)^2¡¡\right.  \nonumber \\
&&¡¡\left.
-2\frac{\tilde \beta }{\Lambda \mu}  \left ((k-f)^2
-\frac{(k-f)^3}{\Lambda r^2} \right )
+\frac{\tilde \beta ^2}{\Lambda^2 \mu^2}\frac{(k-f)^4}{\Lambda^2 r^4}
\right ),
\end{eqnarray}
while for the IR modified Ho\v{r}ava-Lifshtiz theory, we have,
from (\ref{4eq2}),
\begin{equation}
M(r) =\frac{\kappa^2 \mu^2 \Omega}{16 \ r}\left( \Lambda^2 r^4 +2
(\nu-\Lambda) r^2 (1-f) + (1-f)^2\right).
\end{equation}
It is easy to show that, when the matter field is absent, these two
generalized Misner-Sharp energies are conserved, while when the matter
field appears, their variation satisfies (\ref{5eq4}). At the black
hole horizon, the variation of the generalized Misner-Sharp energy
obeys (\ref{5eq5}), which is closely related to the first law of
black hole thermodynamics.

\section{Conclusions and Discussions}

The black hole thermodynamics implies that there might exist a deep
connection between thermodynamics and gravity theory, although they
are quite different subjects. Such a connection must be closely
related to the holographic properties of gravity. The holography
is an essential feature of gravity.  In this paper we investigated
the relationship between the first law of thermodynamics and
gravitational field equation at a static, spherically symmetric
black hole horizon in Ho\v{r}ava-Lifshtiz theory with/without
detailed-balance.  It turns out that, as in the cases of Einstein
gravity and Lovelock gravity, the gravitational field equation can
be cast to a form of the first law of thermodynamics at the black
hole horizon. This way we obtained entropy and mass expressions in
terms of black hole horizon, and they are exactly the same as those
resulting from the integration method for black hole entropy and
the Hamiltonian approach for black hole mass~\cite{CCO}.

Note that Ho\v{r}ava-Lifshtiz theory, different from general relativity, is
not fully diffeomorphism invariant and only keeps the
``foliation-preserving" diffeomorphism. Our results on the relation
between the first law of thermodynamics and gravity field equation
in the Ho\v{r}ava-Lifshtiz theory indicate that this relation is a
robust one, and is of some universality. In addition, unlike the
case in general relativity, the first law of black hole mechanics
has not yet been established so far in Ho\v{r}ava-Lifshtiz theory. Our
result is a first step towards that goal.  Furthermore, let us
stress that in the process to derive the entropy and mass of black
holes in Ho\v{r}ava-Lifshtiz theory, we have not employed an explicit
solution of the theory. This is quite different from the previous works
in the literature. This manifests that the relation between the
first law and gravity field equation has a deep implication.

We also defined generalized Misner-Sharp energy for static, spherically
symmetric spacetimes in Ho\v{r}ava-Lifshtiz theory. The
generalized Misner-Sharp energy is conserved in the case without
matter field, and its variation gives the first law of black hole
thermodynamics at the black hole horizon.

Note that we have restricted ourselves to the case with $\lambda=1$
in Sec.~III.  Here let us make a simple discussion of the case with
$\lambda \neq 1$. In this case, the reduced action can be expressed
as~\cite{CCO}
\begin{equation}
I = \frac{\kappa^2\mu^2 \Omega_k}{8(1-3\lambda)} \int dt dr \tilde
N\left \{
\frac{(\lambda-1)}{2}F'^2-\frac{2\lambda}{r}FF'+\frac{(2\lambda-1)}{r^2}F^2\right\},
\end{equation}
where $F(r) =k-\Lambda r^2-f(r)$. Varying the action with respect to
$F$ and $\tilde N$ yields the equations of motion
\begin{eqnarray}
\label{6eq2}
0 &=& \left (\frac{2\lambda}{r} F-(\lambda-1)F'\right )\tilde N' +
(\lambda-1) \left(\frac{2}{r^2}F-F''\right)\tilde N,
\\
\label{6eq3}
0 &=& \frac{(\lambda-1)}{2}F'^2-\frac{2\lambda}{r}FF'+\frac{(2\lambda-1)}{r^2}F^2.
\end{eqnarray}
These equations have the solution with~\cite{LMP,CCO}
\begin{equation}
\label{6eq4}
F(r) = \alpha r^s, \ \ \  \tilde N(r)= \gamma r^{1-2s},
\end{equation}
where $\alpha$ and $\gamma$ are both integration constants and
$$ s = \frac{2\lambda \pm \sqrt{2(3\lambda-1)}}{\lambda-1}.$$
As discussed in the second reference of \cite{CCO}, to have a
well-defined physical quantities and well-behaved asymptotical behavior
for the solution, we have to take the negative branch in $s$ and $s$ is
in the range $s\in [-1,2)$. The temperature of the black hole in this case is
\begin{equation}
\label{6eq6}
T  =\frac{1}{4\pi}\tilde N(r)f'(r)|_{r=r_+}=\frac{\gamma}{4\pi r_+^{2s}}
\left(-\Lambda r_+^2(2-s)-sk \right ).
\end{equation}
We can rewrite Eq.~(\ref{6eq3}) as
\begin{equation}
\label{6eq7}
\frac{2\lambda}{r}(k-\Lambda r^2-f)f' +4\lambda \Lambda (k-\Lambda
r^2 -f)
+\frac{(\lambda-1)}{2} F'^2 +\frac{(2\lambda-1)}{r^2}F^2=0.
\end{equation}
On the black hole horizon where $f(r_+)=0$, the above equation
reduces to
\begin{equation}
\label{6eq8}
\frac{2\lambda}{r_+}(k-\Lambda r^2_+)f' +4\lambda \Lambda (k-\Lambda
r^2_+ ) +\frac{(\lambda-1)}{2} F'^2(r_+) +\frac{(2\lambda-1)}{r^2}F^2(r_+)=0.
\end{equation}
Multiplying Eq.~(\ref{6eq8}) by
$$
\frac{\sqrt{2}\kappa^2\mu^2\Omega_k
 \tilde N(r_+)}{16\lambda \sqrt{3\lambda-1}}dr_+,
$$
and considering the
expression of the temperature (\ref{6eq6}), we find that the first
term in (\ref{6eq8}) can be expressed as $TdS$, where
\begin{equation}
\label{6eq9}
S = \frac{\pi \kappa^2 \mu^2\Omega_k}{\sqrt{2(3\lambda-1) } } \left (k
\ln (\sqrt{-\Lambda}r_+) +\frac{1}{2}(\sqrt{-\Lambda} r_+ )^2 \right )
+S_0,
\end{equation}
where $S_0$ is an integration constant. On the other hand, with
the solution (\ref{6eq4}), the other three terms in (\ref{6eq8})
can be expressed as $-dM$, where $M$ is
\begin{equation}
\label{6eq10}
M=\frac{\sqrt{2}\kappa^2\mu^2\gamma \Omega_k}{16\sqrt{3\lambda-1}}
\frac{(k-\Lambda r_+^2)^2}{r_+^{2s}}.
\end{equation}
Thus we have shown that on the black hole horizon, the equation of motion (\ref{6eq3})
can be expressed as $TdS-dM =0$, where $S$ and $M$ are just the
black hole entropy and mass, as found in the second reference of
\cite{CCO}. Here we would like to mention that unlike the case of
$\lambda=1$, due to the presence of $F^2$ and $F'^2$ in the
equation of motion, we have to use the black hole solution
(\ref{6eq4}) in order to express the equation of motion in the form
of the first law of thermodynamics. In addition, we here have
discussed the case with the vacuum solution.

\section*{Acknowledgments}

We thank T. Padmanabhan for useful correspondences.  RGC is
supported partially by grants from NSFC, China (No. 10525060, No.
108215504 and No. 10975168) and a grant from MSTC, China (No.
2010CB833004). NO was supported in part by the Grant-in-Aid for
Scientific Research Fund of the JSPS No. 20540283, and also by the
Japan-U.K. Research Cooperative Program. This work is completed
during RGC's visit to Kinki University, Japan with the support of
JSPS invitation fund, the warm hospitality extended to him
is greatly appreciated.



\end{document}